\documentclass[letterpaper]{article}
\usepackage{aaai25}  
\usepackage{times}  
\usepackage{helvet}  
\usepackage{courier}  
\usepackage[hyphens]{url}  
\usepackage{graphicx} 
\usepackage[table]{xcolor}

\usepackage{tcolorbox}  
\tcbuselibrary{breakable}   
\usepackage{natbib}  
\usepackage{caption} 
\usepackage{amsmath}
\usepackage{algorithm}
\usepackage{algorithmicx}
\usepackage{algpseudocode}
\usepackage{multirow}
\usepackage{enumitem}
\usepackage{graphicx}
\usepackage{subfigure}
\usepackage{soul}
\usepackage{booktabs}
\usepackage{amssymb} 
\usepackage{pifont} 
\usepackage{makecell}
\usepackage{multirow}

\usepackage{xspace}

\newcommand{\ie}{\emph{i.e.,}\xspace}

\usepackage{newfloat}
\usepackage{listings}
\DeclareCaptionStyle{ruled}{labelfont=normalfont,labelsep=colon,strut=off} 
\floatstyle{ruled}
\newfloat{listing}{tb}{lst}{}
\floatname{listing}{Listing}
\title{LLM-Powered User Simulator for Recommender System}
\author{
    Zijian Zhang\textsuperscript{\rm 1},
    Shuchang Liu\textsuperscript{\rm 2},
    Ziru Liu\textsuperscript{\rm 3},
    Rui Zhong\textsuperscript{\rm 2},
    Qingpeng Cai\textsuperscript{\rm 2} \thanks{Corresponding authors: Qingpeng Cai, Xiangyu Zhao, and Chunxu Zhang.}, \\
    Xiangyu Zhao\textsuperscript{\rm 3 *},
    Chunxu Zhang\textsuperscript{\rm 1 *},
    Qidong Liu\textsuperscript{\rm 3, 4},
    Peng Jiang\textsuperscript{\rm 2}
}
\affiliations{    
    \textsuperscript{\rm 1} Key Laboratory of Symbolic Computation and Knowledge Engineering of Ministry of Education, Jilin University \\
    \textsuperscript{\rm 2} Kuaishou Technology 
    \textsuperscript{\rm 3} City University of Hong Kong
    \textsuperscript{\rm 4} Xi'an Jiaotong University\\
    \{zhangzijian,zhangchunxu\}@jlu.edu.cn,
    \{liushuchang, zhongrui\}@kuaishou.com,
    ziruliu2-c@my.cityu.edu.hk,
    cqpcurry@gmail.com,
    xianzhao@cityu.edu.hk,
    liuqidong@stu.xjtu.edu.cn,
    jp2006@139.com
}

\usepackage{bibentry}

\begin{document}

\maketitle

\begin{abstract}
User simulators can rapidly generate a large volume of timely user behavior data, providing a testing platform for reinforcement learning-based recommender systems, thus accelerating their iteration and optimization. However, prevalent user simulators generally suffer from significant limitations, including the opacity of user preference modeling and the incapability of evaluating simulation accuracy. In this paper, we introduce an LLM-powered user simulator to simulate user engagement with items in an explicit manner, thereby enhancing the efficiency and effectiveness of reinforcement learning-based recommender systems training. Specifically, we identify the explicit logic of user preferences, leverage LLMs to analyze item characteristics and distill user sentiments, and design a logical model to imitate real human engagement. By integrating a statistical model, we further enhance the reliability of the simulation, proposing an ensemble model that synergizes logical and statistical insights for user interaction simulations. Capitalizing on the extensive knowledge and semantic generation capabilities of LLMs, our user simulator faithfully emulates user behaviors and preferences, yielding high-fidelity training data that enrich the training of recommendation algorithms. We establish quantifying and qualifying experiments on five datasets to validate the simulator's effectiveness and stability across various recommendation scenarios. 
\end{abstract}
\begin{links}
\link{Code}{https://github.com/Applied-Machine-Learning-Lab/LLM_User_Simulator}
\end{links}

\section{Introduction}
Reinforcement Learning (RL)-based recommender systems have become increasingly important due to their capability to capture user preferences \cite{zhao2021dear,zhao2023user,liu2023multi, zhang2020deep} and long-term engagement \cite{liu2023linrec,zhao2018deep,zhao2018recommendations,zhao2019deep}.
They require interactive training where agent learns to make decisions by interacting with an environment.
Online data reflects the real-time user feedback and behavioral patterns, which is critical for continuously refining recommender systems and solving real-world problems \cite{li2023automlp,00040H00Z24,ZhangSHWK23,zhang2024m3oe}.
However, due to the difficulty in obtaining online user interaction data, high collection costs, and user privacy concerns \cite{rl4rs}, effectively simulating user interaction behavior has become an urgent problem to be solved. 
User simulators can quickly generate user behavior data, thus accelerating the evaluation process, and they guarantee user privacy without collection and use of real user data \cite{zhao2021usersim}.

Despite the promising progress of user simulators for recommender systems \cite{rl4rs,recsim,suber,zhangan,llm_sim1}, existing research has two principal deficiencies. 
Firstly, current simulators fail to explicitly model user preferences, which is a critical function for accurately predicting user choices. 
VirtualTaobao \cite{shi2019virtual} leverages GAN to simulate user interaction distribution and
KuaiSim \cite{zhao2023kuaisim} employs an offline trained transformer to emulate user's responses to recommendations. 
Secondly, there is an absence of an efficient evaluative framework to assess the fidelity of simulated interactions with real user behavior. 
Consequently, there is an urgent need for the advancement of user simulators that can operate with higher degrees of fidelity and transparency in replicating the complex dynamics of user-system interactions.

\begin{figure}[!t]
\setlength{\abovecaptionskip}{-1mm}
\setlength{\belowcaptionskip}{-6mm}
{\subfigure{\includegraphics[width=1\linewidth]{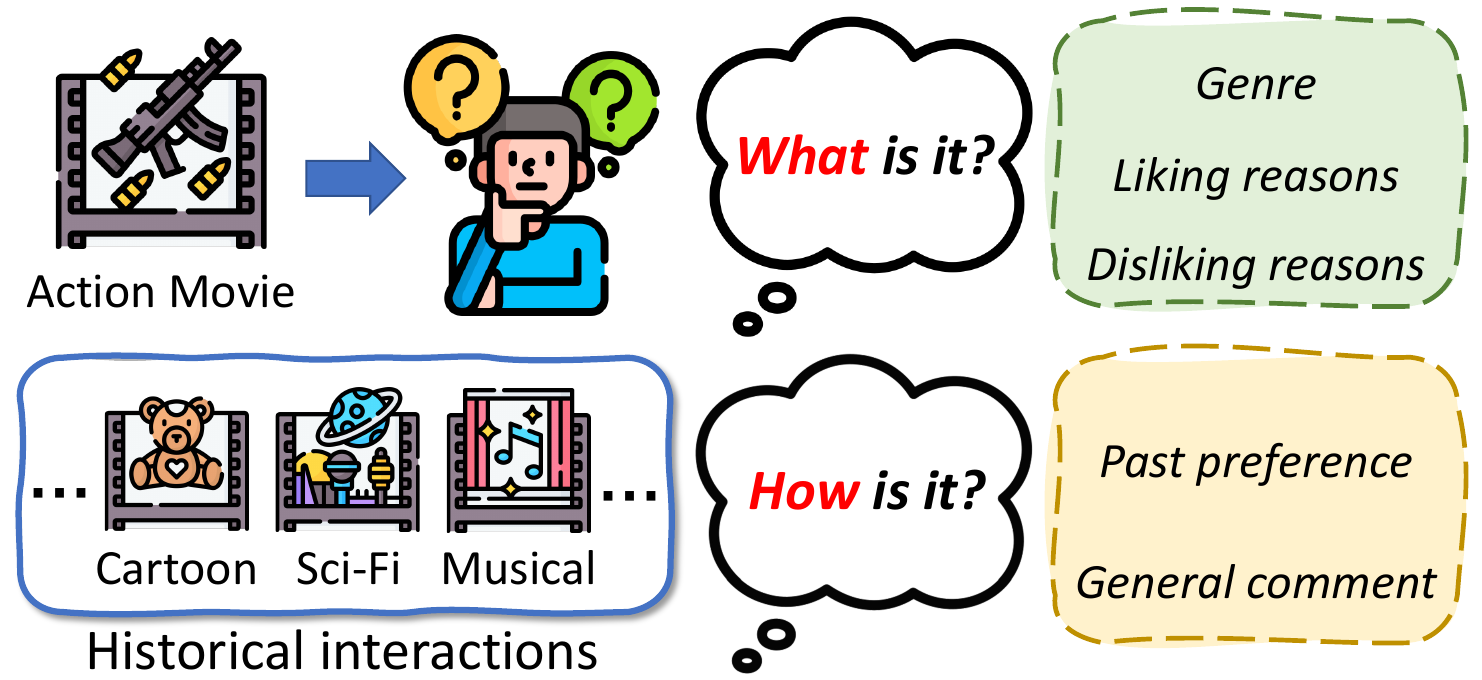}}}

    \caption{
    The user interaction logic for recommended items includes two problems:  understanding what it is and how it is, given the historical interactions.
    }
    \label{fig:pre}
\end{figure}

Recently, Large Language Models (LLMs) have demonstrated remarkable effectiveness in maintaining common knowledge and inferential capabilities across a spectrum of fields \cite{jia2023mill,fu2023unified,wang2024rethinking, wang2024large,liu2024large,liu2024largelanguagemodelenhanced}. 
This prowess positions LLMs as a promising avenue for simulating user behavior in recommendation systems. 
SUBER \cite{suber} harnesses the semantic comprehension capabilities of LLMs to directly infer user engagement with items.
Agent4Rec \cite{zhangan} equips user simulation by integrating LLM agents. It supports a wide range of behaviors, including taste-driven actions, such as viewing and rating items, as well as emotion-driven actions, including exiting the system and feedback through comments.
However, the application of LLMs in user simulation is not without drawbacks. 
Primarily, the computational and time cost of calling LLM for simulations is substantial, posing a significant barrier to the training of the recommendation system \cite{suber, llm_sim1}. 
What's more, an overreliance on LLMs also introduces the risk of hallucination, where the model generates factually incorrect or misleading inferences \cite{hallu1,hallu2}. 
These challenges must be navigated carefully to harness LLMs in crafting user simulators that are both efficient and grounded in reality.


In this paper, we first figure out a fundamental logic governing user interactions with recommended items, as illustrated in Figure \ref{fig:pre}.
Taking movie recommendations as an example, a user's engagement with a recommended action movie begins with recognizing \textbf{what} the movie is, \ie identifying its genre and characteristics that might lead to a liking or disliking sentiment.
This recognition is followed by a deeper inquiry into \textbf{how} the item aligns with the user's tastes, \ie analyzing his past preference to this genre, and considering the general comments to this movie.
Based on item characteristics and the user's preference information, our goal is to explicitly utilize the fundamental logic of user interactions and predict their behavior towards an item in a transparent and understandable manner.


It is nontrivial to achieve this goal, and several main challenges must be conquered. 
First, \textbf{how to depict the user preference explicitly?} 
Translating user preferences into a clear and understandable model is inherently difficult. Existing deep collaborative filtering models with trainable embedding are hardly explainable, let alone indicating user preference.
To address this, we propose leveraging LLM to analyze and interpret item characteristics from various angles, including genre, potential likes and dislikes, and user sentiments. 
By harnessing the rich analytical capabilities of LLM, we develop a logical model that captures the nuanced decision-making processes underlying user interactions.

Second, \textbf{how to alleviate computational cost and the hallucination issue when using LLM? }
The computational demands of LLM and the risk of hallucination present substantial hurdles in system reliability \cite{hallu1}. 
Instead of inferring the user interaction directly based on LLM \cite{zhangan,suber}, 
we use LLM to distill the reasons a user might like or dislike an item, condensing and filtering this reasoning into concise keywords that minimize the impact of outliers. 
In addition, we complement this with a statistical model, \ie a trained sequential model, that provides a regularizing effect on the predictions.
Our ensemble model integrates the strengths of both logical reasoning and statistical learning, enhancing the reliability and efficiency of user interaction simulations synergistically.

Third, \textbf{how to evaluate the user simulator? }
Lacking a standardized metric for evaluation, we face the challenge of assessing the effectiveness of our user simulator. 
To tackle this, we establish a series of experiments that span a wide range of applications, \ie five public datasets encompassing POI, music, movie, game, and anime recommendation, to ensure that our tests are generalizable.
By training reinforcement learning algorithms within these domains and comparing their performance, we aim to validate the simulator's ability to provide meaningful insights into user behavior within recommendation systems.

The main contributions of this paper are as follows:
\begin{itemize}
    \item We identify the intrinsic logic governing user engagement with items in recommendation, and propose a logical model that explicitly infers user interaction.
    \item We construct an ensemble model consisting of rule-based logical and data-driven statistical models to imitate human interaction, maintaining consistency and reducing the likelihood of erroneous inferences.
    \item We conduct both qualifying and quantifying experiments on five benchmark datasets that span a variety of application fields to verify the efficacy of the proposed method.
\end{itemize}

\section{Methodology}

\begin{figure*}[!t]
\centering
\setlength{\abovecaptionskip}{-2mm}
\setlength{\belowcaptionskip}{-4mm}
{\subfigure{\includegraphics[width=1\linewidth]{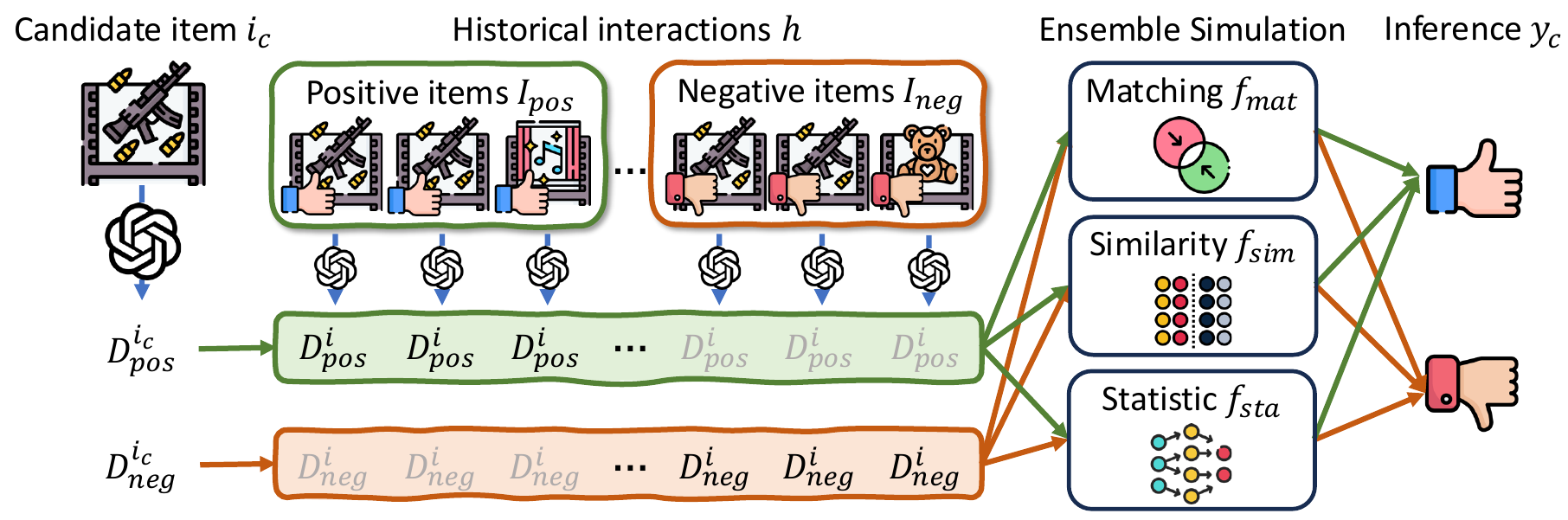}}}

    \caption{User interaction simulation pipeline. We use an LLM to generate preference keywords for items. 
    For a candidate item $i_c$, we separate user historical interactions $h$ into positive and item sets $I_{pos}$ and $I_{neg}$, and apply three base models, \ie two logical models $f_{mat}$, $f_{sim}$, and one statistical model $f_{sta}$, to evaluate $i_c$'s pros and cons, \ie $D_{pos}^{i_c}$ and $D_{neg}^{i_c}$, against the historical item pros and cons.
    We ensemble the three base models' output as the final interaction inference $y_c$.
    }
    \label{fig:framework}
\end{figure*}

\subsection{Preliminaries}
\textit{\textbf{User Simulator.}}
User simulator in recommendation aims to infer the engagement as a real human user with candidate item $i_c$ offered by recommender systems. 
It can be formulated as $P(y_c|h, i_c)$, where $y_c$ represents the interaction given by the user simulator, such as rating, buy, and dislike. In this paper, we consider like and dislike for clear descriptions.
$h=\{(i_1, y_1),\dots, (i_n, y_n)\}$ is the user's historical interactions, consisting of his past responses of $n$ items.

\subsection{Framework Overview}
In this section, we implement the user engagement process with recommended items.
The workflow of our system is depicted in Figure \ref{fig:framework}. The sequence of operations proceeds from left to right, beginning with an analysis of the potential item, followed by a comparison with the user's past interactions, culminating in a reasoned inference.

To express the logic of user interactions in a clear way, we use LLM to examine items and produce informative descriptions. We extract keywords that indicate why items might be liked or disliked based on both the item's features and user reviews.
We also create an ensemble model consisting of three base models. This model suite evaluates the user's past preferences and the current item to produce an inference result. 
By combining these models, we aim to deliver a more precise and dependable assessment.



\subsection{Objective Item Description Collection}
To discern \textbf{what} the item is, we initially leverage the items' factual descriptions.
Our approach is to pinpoint the category $D_{cate}$ and delineate objective reasons for liking or disliking the item, \ie $D_{pos}$ and $D_{neg}$.
This objective item description analysis contributes to a comprehensive understanding of the item's attributes and potential user reactions.

For item categorization $D_{cate}$, we extract textual genre descriptions from the raw data, filtering out the most and least frequent categories. This process ensures that each item is assigned a final category that balances specificity and relevance, avoiding both over-generalization and excessive fragmentation of category representation.

To attain the potential reasons behind user preferences for items $D_{pos}$ and $D_{neg}$, we start with the fundamental details such as the item's name, category, and attributes. Employing an LLM, we generate objective descriptions that encompass the possible factors influencing a user's affinity or aversion towards the item.
Take the movie recommendation as example again, we craft a prompt template $T_{obj}$ that integrates the available information. $T_{obj}$ is shown as follows, where \textcolor{blue}{NAME}, \textcolor{blue}{ATTRIBUTES}, and \textcolor{blue}{CATEGORIES} will be filled with item information from raw data, and the response will fill the placeholder in \textcolor{red}{red text}:

\begin{tcolorbox}[
        colframe=gray,
        width=1\linewidth,
        arc=1mm, 
        auto outer arc,
        title={Objective Description Prompt Template $T_{obj}$},
        breakable,]
        Given a Movie named \textcolor{blue}{NAME} and its characteristics, I need you to provide pros and cons, corresponding evidence and keywords from a customer perspective. Movie has the following attributes: \textcolor{blue}{ATTRIBUTES}. Categories of the movie include \textcolor{blue}{CATEGORIES}. Reasons, keywords, and evidence should be concise and reasonable. Keywords should appear positive or negative. Evidence should refer to the information given. Strictly follow the reply format, fill [], do not say anything else:
        Pros 1: [\textcolor{red}{pros 1}]
        Evidence: [\textcolor{red}{evidence of pros 1}]
        Keywords: [\textcolor{red}{keywords of pros 1}]
        Pros 2: [\textcolor{red}{pros 2}]
        Evidence: [\textcolor{red}{evidence of pros 2}]
        Keywords: [\textcolor{red}{keywords of pros 2}]
        Pros 3: [\textcolor{red}{pros 3}]
        Evidence: [\textcolor{red}{evidence of pros 3}]
        Keywords: [\textcolor{red}{keywords of pros 3}]
        Cons 1: [\textcolor{red}{cons 1}]
        Evidence: [\textcolor{red}{evidence of cons 1}]
        Keywords: [\textcolor{red}{keywords of cons 1}]
        Cons 2: [\textcolor{red}{cons 2}]
        Evidence: [\textcolor{red}{evidence of cons 2}]
        Keywords: [\textcolor{red}{keywords of cons 2}]
        Cons 3: [\textcolor{red}{cons 3}]
        Evidence: [\textcolor{red}{evidence of cons 3}]
        Keywords: [\textcolor{red}{keywords of cons 3}]
        
\end{tcolorbox}

Given the LLM's response according to the form of $T_{obj}$, we extract the keywords that represent positive aspects as $D_{pos}^{obj}$ and those that signify negative aspects as $D_{neg}^{obj}$. 
Consequently, $D_{pos}^{obj}$ and $D_{neg}^{obj}$ emerge as curated lists of keywords, effectively encapsulating the essence of users' potential attitudes towards items. 
These keyword sets serve as concise, informative indicators, reflecting the likely inclinations of user preferences or aversions in a distilled form.

When harnessing the power of LLMs to generate reasons for potential user likes and dislikes, we integrate the Chain of Thoughts (CoT) \cite{cot1,cot2} approach to bolster the precision and reliability of the LLM's output. 
In particular, we guide the LLM to first identify the concrete reasons behind the likes and dislikes using the provided keywords and then summarize the reasons into concise keywords. Finally, we require LLM to substantiate these reasons with evidence drawn from the information provided. 
This process helps circumvent the risk of hallucination inherent in LLMs, ensuring that the generated reasons are well-founded and relevant.

This prompt is engineered to inspire a detailed analysis from the LLM regarding the item's factual characteristics. It encourages the model to consider the full spectrum of attributes from the most appealing features to the possible shortcomings. 
By demanding a logical sequence of identification and justification, the LLM's output is not only informative but also transparent in its reasoning, aligning closely with how human users might evaluate the item.




\subsection{Subjective Item Description Collection}
To comprehend \textbf{how} a user views an item, public opinions can greatly influence the decision-making process. 
For example, a considerable proportion of users tend to choose mainstream options.
Hence, we examine the comments to find keywords that indicate liking or disliking sentiments. 

To accomplish this, we devise a prompt template $T_{sub}$ that encompasses the item's descriptions and the user's feedback.
We ensure that the LLM is instructed to generate reasons for a positive rating when the user provides one, and vice versa for negative ratings. 
For instance, when presenting a prompt for a positive rating, we use ``Pros''; we use ``Cons'' for a negative rating.
The template $T_{sub}$ with positive rating is illustrated as follows, with placeholders for \textcolor{blue}{RATING}, \textcolor{blue}{COMMENTS}, \textcolor{blue}{NAME}, \textcolor{blue}{ATTRIBUTES}, and \textcolor{blue}{CATEGORIES} to be filled with actual data. 
The LLM's response will answer the query indicated by the placeholder in \textcolor{red}{red text}. 

\begin{tcolorbox}[
        colframe=gray,
        width=1\linewidth,
        arc=1mm, 
        auto outer arc,
        title={Subjective Description Prompt Template $T_{sub}$},
        breakable,]
        A customer rates \textcolor{blue}{RATING} to the movie with comments: \textcolor{blue}{COMMENTS} The information of this movie is: name: \textcolor{blue}{NAME}, attributes: \textcolor{blue}{ATTRIBUTES}, categories: \textcolor{blue}{CATEGORIES}.
        I need you to provide {Pros}, corresponding evidence and keywords of the rating based on given information. Evidence should refer to the information given. Strictly follow the reply format, fill [], do not say anything else: 
        {Pros} 1: [\textcolor{red}{{pros} 1}] 
        Keywords: [\textcolor{red}{keywords of {pros} 1}] 
        Evidence: [\textcolor{red}{evidence of {pros} 1}] 
        {Pros} 2: [\textcolor{red}{{pros} 2}] 
        Keywords: [\textcolor{red}{keywords of {pros} 2}] 
        Evidence: [\textcolor{red}{evidence of {pros} 2}] 
        {Pros} 3: [\textcolor{red}{{pros} 3}] 
        Keywords: [\textcolor{red}{keywords of {pros} 3}]
        Evidence: [\textcolor{red}{evidence of {pros} 3}]
        
        
\end{tcolorbox}

Similar to our approach with $T_{obj}$, we prompt the LLM to list the pros, their associated keywords, and supporting evidence, aiming for an output that is both reliable and well-substantiated. 
From the LLM's response, we extract the keywords that indicate positive aspects as $D_{pos}^{sub}$ and those indicating negative aspects as $D_{neg}^{sub}$.

Upon obtaining the lists of keywords that represent likes and dislikes from both the objective and subjective standpoints, we combine these lists to form comprehensive sets: $D_{pos}=D_{pos}^{obj} \cup D_{pos}^{sub}$ and $D_{neg}=D_{neg}^{obj} \cup D_{neg}^{sub}$. 
By merging these keyword lists, we develop a holistic view of the potential inclinations users may have towards items. This synthesis allows us to better anticipate and understand the diverse motivations that drive user preferences and behaviors in the context of recommendations.

We then apply a filtering process to refine these keywords, excluding those that are either too common or too rare. This ensures that the keywords we retain are both informative and concise, effectively capturing the essence of item factual information and real users' subjective opinions. 

In summary, we have attained informative textual descriptions for each item, including category $D_{cate}$, a set of keywords reflecting reasons for liking $D_{pos}$, and disliking keywords $D_{neg}$.
This comprehensive characterization equips our user simulation with a deeper understanding of the items' attributes and the potential reactions from users.


\subsection{Logical Model}
Following the explicit user interaction logic, we design a logical model to simulate user engagement with the recommended candidate item. 
To decide whether to like or dislike a candidate item, users consider the characteristics of historical liking items and the potential reasons for liking candidate items and compare them with the characteristics of historical disliking items and the potential reasons for disliking candidate items.
In this subsection, we devise two types of logical models leveraging the distilled information of items, \ie keywords matching model and similarity calculation model.
\paragraph{Keywords Matching Model}
Given user's historical interacted item list $h=\{(i_k, y_k)\}_{k=1}^n$ containing indicative features, we design a keywords matching model, which concentrates on the direct matching of textual keywords.
Firstly, to retrieve user's preference explicitly, we extract historical interacted items with the same category of $i_c$, noted as $h_C:=\{(i_k, y_k)\}_{k=1}^C$ with $C$ items.
When there are no historical items with the same category, we set $h_C=h$.
Then, we extract item set from $h_C$ with positive rating $y_k=1$ as historical liking items $I_{pos}:=\{i_k\}_{k=1}^{C_{pos}}$ and item set with negative rating $y_k=0$ as historical disliking items $I_{neg}:=\{i_k\}_{k=1}^{C_{neg}}$, where $C = C_{pos} + C_{neg}$.

Given the user historical preference $I_{pos}$ and $I_{neg}$, we design a keywords matching model $f_{mat}(I_{pos}, I_{neg}, i_c)$ to infer the user's inclination towards liking or disliking candidate item $i_c$. 
To implement this, we calculate the number of matched keywords of historical liking reasons, \ie $D_{pos}$ of items in $I_{pos}$, and potential liking reasons for the candidate item, \ie $D_{pos}$ of $i_c$.
Similarly, we calculate the alignment of historical disliking reasons, \ie $D_{neg}$ of items in $I_{neg}$ and potential disliking reasons of candidate item, \ie $D_{neg}$ of $i_c$.
Denote the keywords for positive and negative reasons of item $i$ with $D_{pos}^i$ and $D_{neg}^i$, respectively. The number of matched keywords for positive and negative attitudes $\alpha_{pos}$ and $\alpha_{neg}$ can be formulated as:
\begin{equation}\label{Equ:match_pos}
    \alpha_{pos} = \sum_{i\in I_{pos}}|D_{pos}^{i_c} \cap D_{pos}^i|
\end{equation}
\begin{equation}\label{Equ:match_neg}
    \alpha_{neg} = \sum_{i\in I_{neg}}|D_{neg}^{i_c} \cap D_{neg}^i|
\end{equation}

$\alpha_{pos}$ and $\alpha_{neg}$ capture the degree of overlap between the candidate item's potential reasons for liking/disliking and the user's historical preferences, quantifying the user's potential inclination towards the item.
The keywords matching model can be represented by Eq. \ref{Equ:matching}:
\begin{equation}\label{Equ:matching}
    f_{mat}(I_{pos}, I_{neg}, i_c) = \left\{
    \begin{aligned}
    &1 &if& &\alpha_{pos} > \alpha_{neg},\\
    &rand\{0, 1\} &if& &\alpha_{pos} = \alpha_{neg},\\
    &0 &if& &\alpha_{pos} < \alpha_{neg}.
    \end{aligned}
    \right.
\end{equation}


\paragraph{Similarity Calculation Model}
To enhance the underlying interaction logic simulation process beyond mere keyword matching, we introduce similarity calculation model $f_{sim}(I_{pos}, I_{neg}, I_c)$ that leverages embedding representations for a nuanced understanding of user preferences.
Akin to the keywords matching model, this model focuses on the relationship among items within the same category $h_C$ and its respective positive and negative subsets, \ie $I_{pos}$ and $I_{neg}$.
This categorical analysis allows us to discern patterns and preferences that are specific to the category, thereby enhancing the precision of our user inclination inference.

Initially, we employ a representative backbone model to transform keywords into embeddings in the semantic space. 
Specifically, we harness the capabilities of BERT \cite{bert} to produce embeddings that capture the semantic richness of each keyword set, calculating $E_{pos}$ as the average pooling of the embeddings of the elements in $D_{pos}$, and similarly, $E_{neg}$ as the average pooling of the embeddings in $D_{neg}$, as shown in Eqs. \ref{Equ:sim_emb_pos} and \ref{Equ:sim_emb_neg}:
\begin{equation}\label{Equ:sim_emb_pos}
    E_{pos} = AvePool(\{Bert(d)|d \in D_{pos}\})    
\end{equation}
\begin{equation}\label{Equ:sim_emb_neg}
    E_{neg} = AvePool(\{Bert(d)|d \in D_{neg}\})
\end{equation}

We then proceed to assess the closeness of the candidate item's pros and cons to those of historically liked or disliked items. 
We calculate similarities between the keywords embedding of the user's positively rated items, \ie $E_{pos}$ of items in $I_{pos}$, and embedding of pros from candidate item, \ie $E_{pos}$ of ${i_c}$.
Meanwhile, we calculate similarities between the keywords embedding of the user's negatively rated items, \ie $E_{neg}$ of items in $I_{neg}$, and the cons from candidate item, \ie $E_{neg}$ of ${i_c}$. 
The similarity between keywords for positive and negative attitudes $\beta_{pos}$ and $\beta_{neg}$ can be mathematically presented as follows, where $\phi$ represents similarity metric, and we use cosine similarity.

\begin{equation}\label{Equ:sim_pos}
    \beta_{pos} = Max\{\phi(E_{pos}^{i_c}, E_{pos}^i)|i \in I_{pos}\}
\end{equation}
\begin{equation}\label{Equ:sim_neg}
    \beta_{neg} = Max\{\phi(E_{neg}^{i_c}, E_{neg}^i)|i \in I_{neg}\}
\end{equation}

This comparison with the cosine similarity metric provides a quantifiable and interpretable measure of how the candidate item aligns with the user's historical preferences. The similarity calculation model can be formulated as Eq. \ref{Equ:similarity}:
\begin{equation}\label{Equ:similarity}
    f_{sim}(I_{pos}, I_{neg}, i_c) = \left\{
    \begin{aligned}
    &1 &if& &\beta_{pos} > \beta_{neg},\\
    &rand\{0, 1\} &if& &\beta_{pos} = \beta_{neg},\\
    &0 &if& &\beta_{pos} < \beta_{neg}.
    \end{aligned}
    \right.
\end{equation}


\begin{table}[t]
\renewcommand{\arraystretch}{1.1}
\centering
\setlength{\abovecaptionskip}{0.1mm}
\caption{Dataset statistics. 
}
\label{dataset}
\resizebox{0.5\textwidth}{!}{
\begin{tabular}{ccccc}
\toprule[1pt]
\textbf{Dataset} & \#Users & \#Items & \#Instances & Sparsity\\
\midrule
\textbf{Yelp} & 15,081 & 10,186 & 228,000 & 99.85\% \\
\textbf{Amazon Music} & 125,627 & 65,019 & 162,261 & 99.99\%  \\
\textbf{Amazon Games} & 30,195 & 23,096 & 165,571 & 99.98\%  \\
\textbf{Amazon Movie} & 24,191 & 49,154 & 972,536 & 99.92\%  \\
\textbf{Anime} & 24,859 & 11,188 & 6,111,860 & 97.80\%  \\

\bottomrule[1pt]
\end{tabular}}
\end{table}

\subsection{Statistic Model}
Beyond the two logical models, we augment our user simulator with a statistical model to enhance the reliability of the generated user inferences. 
To achieve this, we employ a deep model, SASRec \cite{sasrec}, and train on the user's historical interaction data. 
Specifically, we pretrain the statistical model $f_{sta}(h, i_c)$ with the dataset. 
Subsequently, we load the pretrained model parameter to infer the engagement with the candidate item.

This approach introduces the power of traditional statistical model, trained on user historical interaction data, to predict the user's inclination towards a candidate item. 
It enhances the reliability of the user simulator by integrating statistical learning with logical reasoning.

\subsection{Ensemble User Simulator}
Given the established two logical models, \ie keywords matching model $f_{mat}$ and similarity calculation model $f_{sim}$, and statistic model $f_{sta}$, we construct an ensemble model to synergize the user interaction simulation performance.

By integrating the strengths of both logical reasoning and statistical learning, our ensemble model offers a comprehensive and robust framework for simulating user preferences and behaviors in recommendation scenarios.

Next we introduce the training pipeline with reinforcement learning-based recomender system in this subsection.

\subsection{MDP Formulation}
In reinforcement learning-based recommender system training, the sequential item interactions can be formulated by a Markov Decision Process (MDP) \cite{mdp}.
In this pipeline, recommender system serves as an agent, user interaction and preference are the environments, recommendation of item is action, user's response towards recommendation is the reward signal.

\begin{itemize}
    \item \textbf{State} ($s \in \mathcal{S}$): the set of all possible states of the environment, including user profile, historical interactions $h$, and current context including $I_{pos}$ and $I_{neg}$.
    \item \textbf{Action} ($i_c \in \mathcal{A}$): the set of all possible actions the recommender system can take, where an action represents one recommended item $i_c$.
    \item \textbf{Transition Probabilities} ($\mathcal{P}(s'|s,i_c)$): the probabilities of transitioning to a new state $s'$ given the current state $s$ and action $i_c$ from recommender system.
    \item \textbf{Reward Function} ($\mathcal{R}(s,i_c,s')$): assigns a numerical reward to each transition from state $s$ to $s'$ by action $i_c$. 
We craft an ensemble model to serve as the user simulator, and the reward function is determined by the consensus reached among three constituent base models, which could be formulated as Eq. \ref{Equ:reward}:
\begin{equation}\label{Equ:reward}
    \mathcal{R}(s,i_c,s') = \left\{
    \begin{aligned}
    1& &if& &f_{mat}+f_{sim}+f_{sta} \geq 2,\\
    0& &if& &f_{mat}+f_{sim}+f_{sta} < 2.
    \end{aligned}
    \right.
\end{equation}
    \item \textbf{Discount Factor} $\gamma$: A number between 0 and 1 used to discount future rewards.
\end{itemize}

\begin{table}[!ht]
    \centering
    \caption{Overall performance. A. Rwd and T. Rwd represent average and total rewards, respectively. Liking\% is the liking items ratio in the top-10 recommendations.
    }
    \label{tab:overall}
    \resizebox{0.5\textwidth}{!}{
    \begin{tabular}{c|c|cccc}
\toprule[1pt]
        \textbf{Dataset} & \textbf{Metric} & PPO & TRPO & A2C & DQN \\ \midrule
        \multirow{3}{*}{\textbf{Yelp}} & \textbf{A. Rwd}$\uparrow$ & 9.97 & {13.45} & {24.15} & \textbf{27.56} \\ 
         & \textbf{T. Rwd}$\uparrow$ & {141.57} & {157.42} & {267.60} & \textbf{330.98} \\ 
         & \textbf{Liking\%}$\uparrow$ & 34.59 & 40.07 & 48.35 & \textbf{49.43} \\ \midrule 
        \multirow{3}{*}{\makecell{\textbf{Amazon} \\ \textbf{Music}}} & \textbf{A. Rwd}$\uparrow$ & 10.49 & {11.31} & 13.45 & \textbf{16.70} \\ 
         & \textbf{T. Rwd}$\uparrow$ & {129.03} & 140.15 & 141.03 & \textbf{181.42} \\ 
         & \textbf{Liking\%}$\uparrow$ & 29.30 & 32.46 & 29.54 & \textbf{33.18} \\ \midrule
        \multirow{3}{*}{\makecell{\textbf{Amazon} \\ \textbf{Games}}} & \textbf{A. Rwd}$\uparrow$ & 18.72 & {21.35} & \textbf{27.56} & {26.43} \\ 
         & \textbf{T. Rwd}$\uparrow$ & 208.43 & 242.26 & \textbf{317.56} & {269.02} \\ 
         & \textbf{Liking\%}$\uparrow$ & 33.15 & 37.64 & \textbf{43.52} & 40.73 \\ \midrule
        \multirow{3}{*}{\makecell{\textbf{Amazon} \\ \textbf{Movie}}} & \textbf{A. Rwd}$\uparrow$ & 29.42 & {27.47} & 31.72 & \textbf{38.60} \\ 
         & \textbf{T. Rwd}$\uparrow$ & {310.69} & 301.40 & 354.34 & \textbf{416.18} \\ 
         & \textbf{Liking\%}$\uparrow$ & 38.59 & 36.70 & 42.37 & \textbf{44.50} \\ \midrule
        \multirow{3}{*}{\textbf{Anime}} & \textbf{A. Rwd}$\uparrow$ & 14.12 & 14.58 & \textbf{21.50} & {18.03} \\ 
         & \textbf{T. Rwd}$\uparrow$ & 155.74 & 163.44 & \textbf{242.95} & 201.94 \\ 
         & \textbf{Liking\%}$\uparrow$ & 25.46 & 24.27 & \textbf{31.52} & 30.67 \\ 
\bottomrule[1pt]
    \end{tabular}
    }
\end{table}

\section{Experiments}

\subsection{Experimental Setup}

To verify the efficacy of the proposed ensemble user simulator, we incorporate datasets from diverse fields: \textbf{Yelp\footnote{\url{https://www.yelp.com/dataset/documentation/main}}} (the state of Missouri), 
\textbf{Amazon\footnote{\url{https://nijianmo.github.io/amazon/index.html}}} 
(Digital Music, Video Games, and Movies), and \textbf{Anime\footnote{\url{https://www.kaggle.com/datasets/CooperUnion/anime-recommendations-database}}}.
Dataset statistics are shown in Table \ref{dataset}. 
We convert ratings into a binary format, where a rating is marked as '1' if it is 3 or higher and '0' for ratings below 3. 
We use ChatGLM-6B\footnote{\url{https://huggingface.co/THUDM/chatglm-6b}} as our LLM.
We employ several representative reinforcement learning algorithms: \textbf{A2C} \cite{a2c}, \textbf{DQN} \cite{dqn}, \textbf{PPO} \cite{ppo}, and \textbf{TRPO} \cite{trpo}. 

\subsection{Benchmark Results}
We provide the results of reinforcement learning algorithms on our user simulator and report the average reward, total reward, and liking ratio. Experimental results are shown in Table \ref{tab:overall}.
The results indicate that DQN consistently outperforms other algorithms, a result that can be attributed to its superior capacity for handling tasks with discrete action spaces. DQN combines Q-learning with deep learning and excels at estimating the expected return for each action. The robust performance of DQN in the simulator is likely bolstered by its use of experience replay and target networks.

Besides, all algorithms exhibit good liking ratio of recommendation, which suggests that the user simulator provides a consistent environment where different algorithms can perform their interactions with the recommended items over a similar timescale.
The user simulator can produce consistent and reliable interaction patterns across different algorithms, showcasing the simulator's reliability in mimicking real user behavior.
It highlights the simulator's strength in replicating genuine user behaviors, which is crucial for accurately assessing algorithmic performance.



\subsection{Case Study}
In this subsection, we delve into case studies to further demonstrate the effectiveness of our user simulator.
We first illustrate the recommendations by the DQN on Yelp in Table \ref{tab:case_rl}.
Due to space limitation, we showcase a selection of 3 historical items ($i^{t-3}$, $i^{t-2}$, and $i^{t-1}$) and 3 RL recommended items ($i_c^t$, $i_c^{t+1}$, and $i_c^{t+2}$) highlighting their notable pros and cons.
We omit cons for positive historical items and pros for negative ones.
The user simulator's detailed inference process on the RL recommendations is presented in Table \ref{tab:case_simulator}.
For instance, for $i_c^t$, the matching cons with $i^{t-1}$ results in a $f_{mat}$ output of 0.
Similarly, $f_{mat}$ for $i_c^{t+2}$ is 1, as its pros align with those of $i_{t-2}$.
The $f_{mat}$ for $i_c^{t+1}$ being 1 is incidental, given the lack of matching pros or cons.

When faced with items featuring new genres, such as $i_c^{t+1}$, the logical model assesses both the matching of pros and cons and the overall similarity to the historical item set, which may somewhat reduce precision. Nonetheless, our statistical model $f_{sta}$ serves as a crucial fallback. 
Its collaborative filtering capabilities ensure that the inference remains accurate and well-informed.

In conclusion, our ensemble user simulator harnesses the advantages of both logical and statistical models to explicitly generate user interactions that reflect user preferences. The logical model ensures transparency and consistency in user engagement, while the statistical model captures the subtleties of user behavior, enhancing the simulator's fidelity and effectiveness in emulating real-world user interactions. 


\begin{table*}[t]
\renewcommand{\arraystretch}{1.1}
\centering
\caption{RL algorithm recommendation on Yelp. Green grids denote the positive aspects of historical items, and orange ones represent the negative aspects, aligning with the framework in Figure \ref{fig:framework}.
}
\label{tab:case_rl}
\resizebox{1\textwidth}{!}{
\begin{tabular}{ccccccc}
\toprule[1pt]
  & \multicolumn{3}{c}{historical items $h$}  & \multicolumn{3}{c}{RL recommendations $i_c$} \\
  \cmidrule(lr){2-4}\cmidrule(lr){5-7}
  & $i^{t-3}$ & $i^{t-2}$ & $i^{t-1}$ & $i_c^t$  & $i_c^{t+1}$  & $i_c^{t+2}$ \\
\midrule
{Name} & City Diner & EI Maguey & Popeyes Kitchen & IHOP & Gooey Cakes & Crusoe's Original \\
{Category} & Restaurants & Restaurants & Restaurants & Restaurants & Bakeries & Restaurants \\
{Pros} & \multicolumn{1}{>{\columncolor[HTML]{ddebd3}}c}{family gathering} & \multicolumn{1}{>{\columncolor[HTML]{ddebd3}}c}{child-friendly} & \multicolumn{1}{>{\columncolor[HTML]{ddebd3}}c}{-} & casual & variety &  child-friendly  \\
{Cons} & \multicolumn{1}{>{\columncolor[HTML]{f8ded0}}c}{-} & \multicolumn{1}{>{\columncolor[HTML]{f8ded0}}c}{-} & \multicolumn{1}{>{\columncolor[HTML]{f8ded0}}c}{loud, crowded} & loud & no in-store dining & no reservations  \\
{Rating} & 1 & 1 & 0 & - & - & -  \\

\bottomrule[1pt]
\end{tabular}}
\end{table*}




\begin{table}[t]
\centering
\caption{User simulator inference on recommendation. 
}
\label{tab:case_simulator}
\scalebox{1}{
\begin{tabular}{ccccc}
\toprule[1pt]
  Recommendation & $f_{mat}$ & $f_{sim}$ & $f_{sta}$ & $R$ \\
\midrule
$i_c^t$ & 0 & 0 & 1 & 0 \\
$i_c^{t+1}$ & 1 & 0 & 1 & 1 \\
$i_c^{t+2}$ & 1 & 1 & 1 & 1 \\

\bottomrule[1pt]
\end{tabular}}
\end{table}



\begin{table}[!t]
\renewcommand{\arraystretch}{1.1}
\centering
\caption{User simulator qualified comparison. 
}
\label{tab:simulator}
\resizebox{0.5\textwidth}{!}{
\begin{tabular}{ccccc}
\toprule[1pt]

\textbf{Simulators} & \textbf{Real dataset} & \textbf{Simulation engine }& \textbf{Evaluation} \\
\midrule
\makecell[c]{RecoGym \\ \cite{recogym}} & \ding{53} & Statistical model & case study  \\
\makecell[c]{RecSim \\ \cite{recsim}} & \ding{53} & Statistical model & case study  \\
\makecell[c]{VirtualTaobao \\ \cite{shi2019virtual}} & \checkmark & GAN & online \\
\makecell[c]{Adversarial \\ \cite{chen2019adversarial}} & \checkmark & GAN & offline \\
\makecell[c]{KuaiSim \\ \cite{zhao2023kuaisim}} & \checkmark & Transformer & offline  \\
\makecell[c]{SUBER \\ \cite{suber}} & \checkmark & LLM & offline \& case study \\
\makecell[c]{Agent4Rec \\ \cite{zhangan}} & \checkmark & LLM & offline \& case study \\

\midrule
Ours & \checkmark & \makecell[c]{LLM-based logical \\ \& statistical model} & offline \& case study  \\

\bottomrule[1pt]
\end{tabular}}
\end{table}

\begin{table}[t]
\renewcommand{\arraystretch}{1.1}
\centering
\caption{User simulator quantified comparison. 
}
\label{tab:simu_comp}
\resizebox{0.4\textwidth}{!}{
\begin{tabular}{ccccc}
\toprule[1pt]
\textbf{Metric}         & \textbf{A. Rwd↑} & \textbf{T. Rwd↑} & \textbf{AUC↑}  & \textbf{Infer Time(s)} \\
\midrule
SUBER       & 23.74             & 297.48              & 0.643          & 2.42                   \\
KuaiSim & 25.35             & 316.46              & 0.658          & 0.53                   \\
Ours                    & \textbf{27.56}    & \textbf{330.98}     & \textbf{0.674} & 0.76  \\
\bottomrule[1pt]
\end{tabular}}
\end{table}

\subsection{User Simulator Comparison}
To clearly position our user simulator with existing user simulators, we present the main features in Table \ref{tab:simulator}.

For RL-based recommender systems, traditional user simulators typically rely on statistical models to generate user inferences. 
our simulator provides a transparent and logical method for inferring user engagement, enhancing transparency and realism.
The reliance on LLMs for inference will inevitably introduce implementation complexity and computational cost, particularly when dealing with large datasets or high-frequency user interactions.
Moreover, the hallucination in LLMs can impede performance, which is an issue that still lacks an effective solution.
Unlike other LLM-based simulators that face issues like computational cost and hallucination during training, our approach utilizes the LLM for offline analysis, avoiding direct calls during the training phase and thus mitigating related challenges.

According to the performance comparison in Table \ref{tab:simu_comp}, our simulator consistently outperforms the state-of-the-art simulators, which proves its precise approximation to user preference. It also achieves competitive efficiency against both LLM- and non-LLM-based simulators. 

\section{Related Works}

\noindent\textbf{Traditional User Simulator}
To bridge the performance gap between offline metrics and online performance of recommender systems, RecoGym \cite{recogym} simulates user behavior in e-commerce and their responses to recommendations. 
Both organic user interactions on e-commerce sites and bandit sessions are incorporated.
RecSim \cite{recsim} provides a customizable environment for testing user interaction hypotheses, allowing for tailored user preferences, latent states, dynamics, and choice models.
Virtual-Taobao \cite{shi2019virtual} uses GANs for realistic customer feature simulation and multi-agent adversarial imitation learning for generalized customer action generation.
Similarly, \citeauthor{chen2019adversarial} \cite{chen2019adversarial} uses GANs to model online interaction rewards, introducing a model-based RL technique that enhances sample efficiency in policy learning.
Kuaisim \cite{zhao2023kuaisim} integrates a transformer model for user responses and sets benchmarks at the request, session, and cross-session levels for comprehensive recommender system evaluation.


\noindent\textbf{LLM-based User Simulator}
Given the successful precedents in related areas \cite{llm1,llm2}, there emerge LLM-based user inference simulations leveraging LLM's outstanding semantic understanding and inferring capability.
With an empirical study on ChatGPT's performance in conversational recommendation using benchmark datasets, \citeauthor{wang2023rethinking} \cite{wang2023rethinking} advocate to consider two types of interaction: attribute-based question answering and free-form chit-chat.
Aiming at simulating search users, USimAgent \cite{zhang2024usimagent} devises LLM agent to fabricate complete search sessions, including querying, clicking, and stopping behaviors, based on specific search tasks.
SUBER \cite{suber} utilizes LLMs as synthetic users within a gym environment, marking progress towards more realistic training environments for recommender systems.
Agent4Rec \cite{zhangan} focuses on developing a simulator that accurately reflects user preferences and social traits. 
It leverages LLMs to create agents, each initialized with a unique user profile that includes tastes and social traits, ensuring agents' behaviors mirror those of real human users.

\section{Conclusion}

This paper presents a novel user simulator for RL-based recommender systems. To address the prevalent issues of opacity and simulation evaluation in current systems in existing user simulators, we introduce an LLM-powered user simulator designed to explicitly model user preferences and engagement. 
Our method identifies explicit logic of user preferences, utilizing LLMs to analyze item characteristics and distill user sentiments. We propose a novel ensemble model that integrates both logical and statistical insights, enhancing the reliability and fidelity of user interaction simulations. 
We conduct comprehensive experiments across five datasets, demonstrating the simulator's effectiveness and stability in various recommendation scenarios.

Currently, this simulator only infers binary interactions of `like' or `dislike'.
Future work will focus on integrating additional interaction signals, such as duration, rating, and retention, to enrich the application of the simulator.

\section{Acknowledgment}

This research was partially supported by Research Impact Fund (No.R1015-23), APRC - CityU New Research Initiatives (No.9610565, Start-up Grant for New Faculty of CityU), CityU - HKIDS Early Career Research Grant (No.9360163), Hong Kong ITC Innovation and Technology Fund Midstream Research Programme for Universities Project (No.ITS/034/22MS), Hong Kong Environmental and Conservation Fund (No. 88/2022), and SIRG - CityU Strategic Interdisciplinary Research Grant (No.7020046), the Fundamental Research Funds for the Central Universities, JLU, and Kuaishou.

\bibliography{aaai25}

\end{document}